\documentclass[a4paper,11pt]{article}
\usepackage[margin=1in]{geometry}
\usepackage[utf8]{inputenc}
\usepackage{graphicx}
\usepackage{setspace}
\usepackage{subcaption}
\usepackage[clean,pdf]{}
\usepackage{float}
\usepackage{amsmath}
\usepackage{amsthm}
\usepackage{amssymb}
\usepackage{epsfig}
\usepackage{xspace}
\usepackage{color}
\usepackage{multicol}
\usepackage{xcolor}

\usepackage{authblk}
\usepackage[labelfont=bf,font={small}]{caption}
\usepackage[backend=bibtex,style=nature]{biblatex}
\bibliography{ref}

\title{Quantifying the rise and fall of scientific fields}
\author[1]{Chakresh Singh}
\author[1]{Emma Barme}
\author[2]{Robert Ward}
\author[1,3]{Liubov Tupikina}
\author[1,*]{Marc Santolini}
\affil[1]{Université de Paris, INSERM U1284, Center for Research and Interdisciplinarity (CRI), F-75006 Paris, France}
\affil[2]{School of Public Policy, Georgia Institute of Technology, Atlanta, GA 30332}
\affil[3]{Nokia Bell labs, France}
\affil[*]{Corresponding author: marc.santolini@cri-paris.org}

\date{}

\begin{document}

\maketitle

\begin{abstract}
 
Science advances by pushing the boundaries of the adjacent possible. While the global scientific enterprise grows at an exponential pace, at the mesoscopic level the exploration and exploitation of research ideas is reflected through the rise and fall of research fields. The empirical literature has largely studied such dynamics on a case-by-case basis, with a focus on explaining how and why communities of knowledge production evolve. Although fields rise and fall on different temporal and population scales, they are generally argued to pass through a common set of evolutionary stages.To understand the social processes that drive these stages beyond case studies, we need a way to quantify and compare different fields on the same terms. In this paper we develop techniques for identifying scale-invariant patterns in the evolution of scientific fields, and demonstrate their usefulness using 1.5 million preprints from the arXiv repository covering $175$ research fields spanning Physics, Mathematics, Computer Science, Quantitative Biology and Quantitative Finance. We show that fields consistently follows a rise and fall pattern captured by a two parameters right-tailed Gumbel temporal distribution. We introduce a field-specific rescaled time and explore the generic properties shared by articles and authors at the creation, adoption, peak, and decay evolutionary phases. We find that the early phase of a field is characterized by the mixing of cognitively distant fields by small teams of interdisciplinary authors, while late phases exhibit the role of specialized, large teams building on the previous works in the field. This method provides foundations to quantitatively explore the generic patterns underlying the evolution of research fields in science, with general implications in innovation studies. 

\end{abstract}

\section{Introduction}

Quantifying the dynamics of scientific fields can help us understand the past and design the future of scientific knowledge production. Several studies have investigated the emergence and evolution of scientific fields, from the discovery of new concepts to their adaptation and modification by the scientific community \cite{frickel2005general,shwed2010temporal,sun2013social,jurgens2018measuring}. In particular, methods ranging from bibliometric studies \cite{bettencourt2008population,dong2017allometric} to network analyses \cite{herrera2010mapping,sun2016mapping,jurgens2018measuring} and natural language processing \cite{balili2020termball,dias2018using} have been implemented on large publication corpora to monitor the propagation of concepts across articles \cite{chavalarias2013phylomemetic,Sun2020} and the social interactions between researchers that are producing them \cite{sun2016mapping,bettencourt2009scientific,bettencourt2015formation}. \\

The definition of research fields often relies on data-driven strategies using self-reported keywords or content analysis. For example, the use of granular author self-reported topic annotations from well-defined classification schemes such as PACS (Physics and Astronomy Classification Scheme), MESH terms and keywords has allowed to construct topic co-occurrence networks and extract clusters corresponding to potential research fields \cite{herrera2010mapping,balili2020termball}. Beyond self-reported annotations, other methods have exploited the citation network between research articles to group articles by relatedness and map the knowledge flow within and across research fields \cite{Sun2020,jurgens2018measuring}, and inferential methods have leveraged Natural Language Processing techniques to automatically identify key topics of research \cite{chavalarias2013phylomemetic,dias2018using,dalle2020understanding} and their relations. These various methods provide clusters of closely related topics corresponding to putative research fields, allowing to monitor how the changing relations between topics and ideas underlie the dynamic evolution and mutual interactions between fields. Beyond topic-centric approaches, other methods have leveraged the interactions between researchers to define research communities with shared research interests. For example, the co-authorship network between researchers has been shown to undergo a topological transition during the emergence of a new field \cite{bettencourt2009scientific,bettencourt2015formation}. Co-authorship relations also influence the individual evolution of research interests and foster the emergence of a consensus in a research community \cite{jia2017quantifying,zeng2019increasing,bonaventura2017advantages}. \\

While fields rise and fall on different temporal and population scales, they are generally argued to pass through a common set of evolutionary stages \cite{kuhn2012structure,frickel2005general}. These stages delineate how diverse actors and behaviors are involved in successive phases. To study these temporal patterns, dynamical models were introduced to characterize the evolution of research fields \cite{scharnhorst2012models, bettencourt2008population} and the spread of innovation \cite{rogers2010diffusion,robertson1967process,katz1963traditions}. Yet, we are still lacking a unified framework to delineate stereotyped stages in the evolution of scientific fields that can be validated over a large number of well-annotated research fields.\\

Here, we address this gap by developing techniques for identifying scale-invariant patterns in the evolution of fields. We demonstrate their usefulness using a large corpus of 1.45 million articles from the arXiv repository with self-reported field tags spanning 175 research fields in Physics, Computer Science, Mathematics, Finance, and Biology. We show that the evolution of fields follows a right-tailed distribution with two parameters characterizing peak location and distribution width. This allows us to collapse the temporal distributions onto a single rise-and-fall curve and delineate different evolutionary stages of the fields: creation, adoption, peak, early decay, and late decay. We then describe the characteristics of articles and authors across these stages. We finish by discussing these results and their implications for further work in science and innovation. 

\begin{figure}[!htb]
    \centering
    \includegraphics[width=\textwidth]{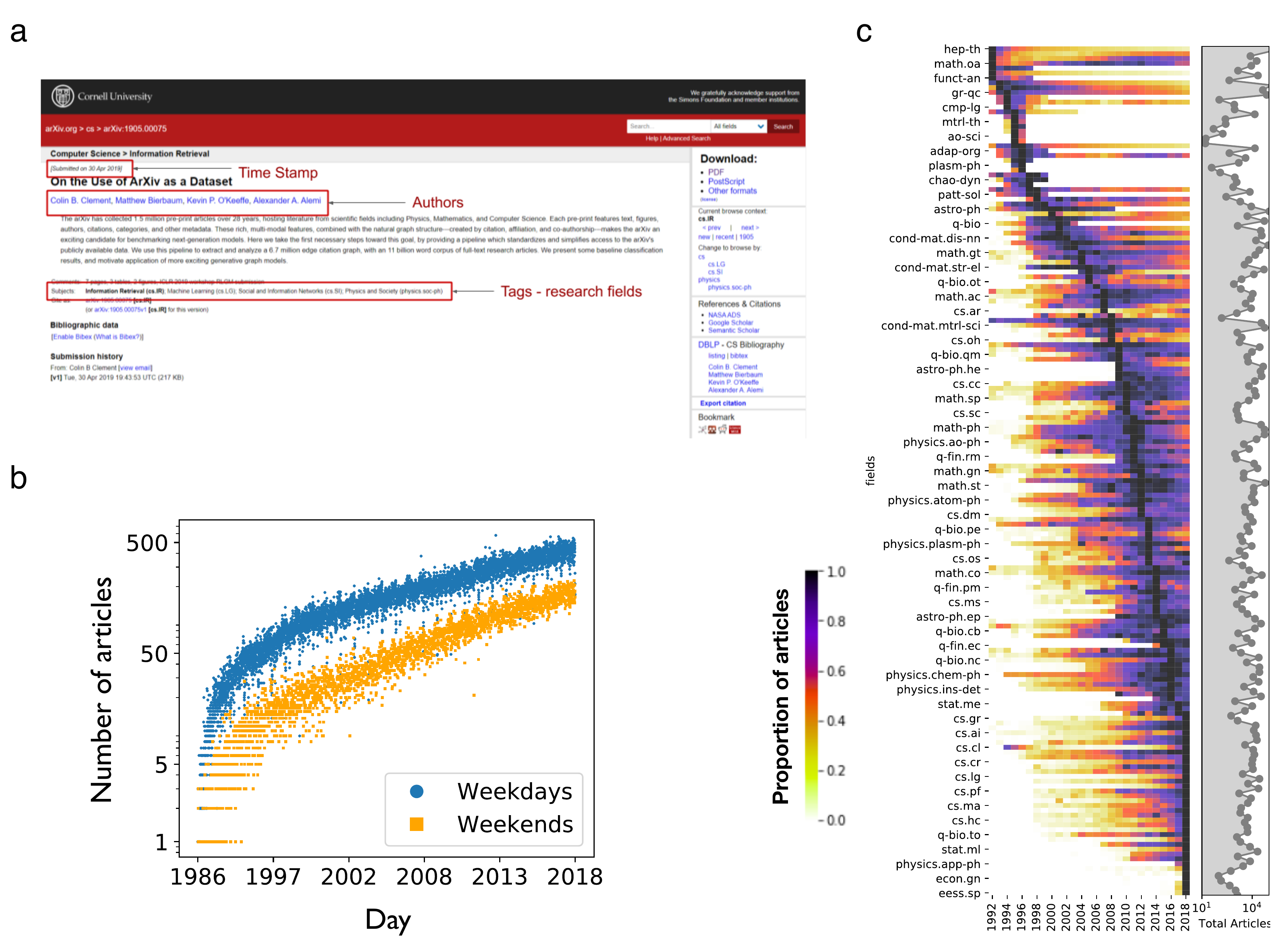}
    \caption{\textbf{a} Example of an article in arXiv, highlighting the metadata extracted using the arXiv API. \textbf{b} The daily number of articles submitted to arXiv since 1986 shows an exponential growth over time, with a doubling period of 6 years. The data also shows strong seasonality with 10 times fewer articles over the weekends. \textbf{c} Heatmap representing the share of articles in each field (rows) over time (columns). Field are identified using the subject tags within articles. The heatmap is row normalized for comparison across fields. Rows are ordered in chronological order of their peak time. The right side panel shows the total number of articles published in each field as horizontal bars.}
    \label{fig1}
\end{figure}

\section{Results}

\subsection{Description of the data}

Since its launch in $1991$, the arXiv repository has become a major venue for community research, gaining considerable importance across the fields of Physics, Mathematics and Computer Science. As an open and free contribution platform, it provides an equal opportunity for publication to researchers globally, and plays a dominant role in the diffusion of knowledge \cite{lariviere2014arxiv} and the evolution of new ideas \cite{sun2020evolution}.\\

When submitting a contribution, authors declare the research fields that the article is contributing to by selecting from a list of subject tags. Here we collected information about authors, date of publication, and research fields of 1,456,403 arXiv articles until 2018 (see Methods section and Fig \ref{fig1}a). The number of articles and authors exhibit an exponential growth over time  with a doubling period of 6 years (see Fig \ref{fig1}b and \ref{figS:num_authors}). To control for this effect, here we focus for each field $i$ on the yearly share of articles $f_{i,y} = n_{i,y} / N_y$, where $n_{i,y}$ is the number of articles published in the considered field at year $y$ and $N_y$ is the total number of articles in arXiv in the same year. We represent the temporal distributions of all fields in Fig \ref{fig1}c by chronological peak time. Over the past 30 years, the research interests have shifted from high-energy physics to computer science and more recently economy.  

\subsection{Quantifying the rise and fall of scientific fields}

Despite differences in overall number of articles and eventual duration, we observe a general rise-and-fall pattern across research fields (Fig \ref{fig1}c), prompting us to explore if a simple model can capture their temporal variation.
Extreme value theory \cite{kotz2000extreme}  predicts that under a broad range of circumstances, temporal processes displaying periods of incubation (such as incubation of ideas) or processes with multiple choice (such as the choice of ideas or research fields) follow skewed right-tailed extreme value distributions. Examples of such processes can be found in diverse areas, for example when modeling the evolution of scientific citations \cite{sinatra2016quantifying} or disease incubation periods \cite{bertrand2017evolutionary, gautreau2008global}. Following these insights, here we use the Gumbel distribution (Eq. \ref{g_pdf}) as an ansatz to model the observed field temporal distributions. Belonging to the general class of extreme value distributions \cite{gumbel1958statistics, kotz2000extreme}, it provides interpretable parameters for the peak location $\alpha$ and distribution width $\beta$ (Fig. \ref{figS:gumble_shapes}). Denoting by $t$ the time since the first article was published in the field, the share of articles $G(t)$ follows Eq. \ref{g_pdf}:

\begin{equation}\label{g_pdf}
    G(t) = \frac{1}{\beta}e^{\frac{-(t-\alpha)}{\beta}}e^{-e^{\frac{-(t-\alpha)}{\beta}}}
\end{equation}

where $\alpha$ is the location parameter and $\beta$ the scale parameter.\\

\begin{figure}[ht!]
    \centering
    \includegraphics[width=\textwidth]{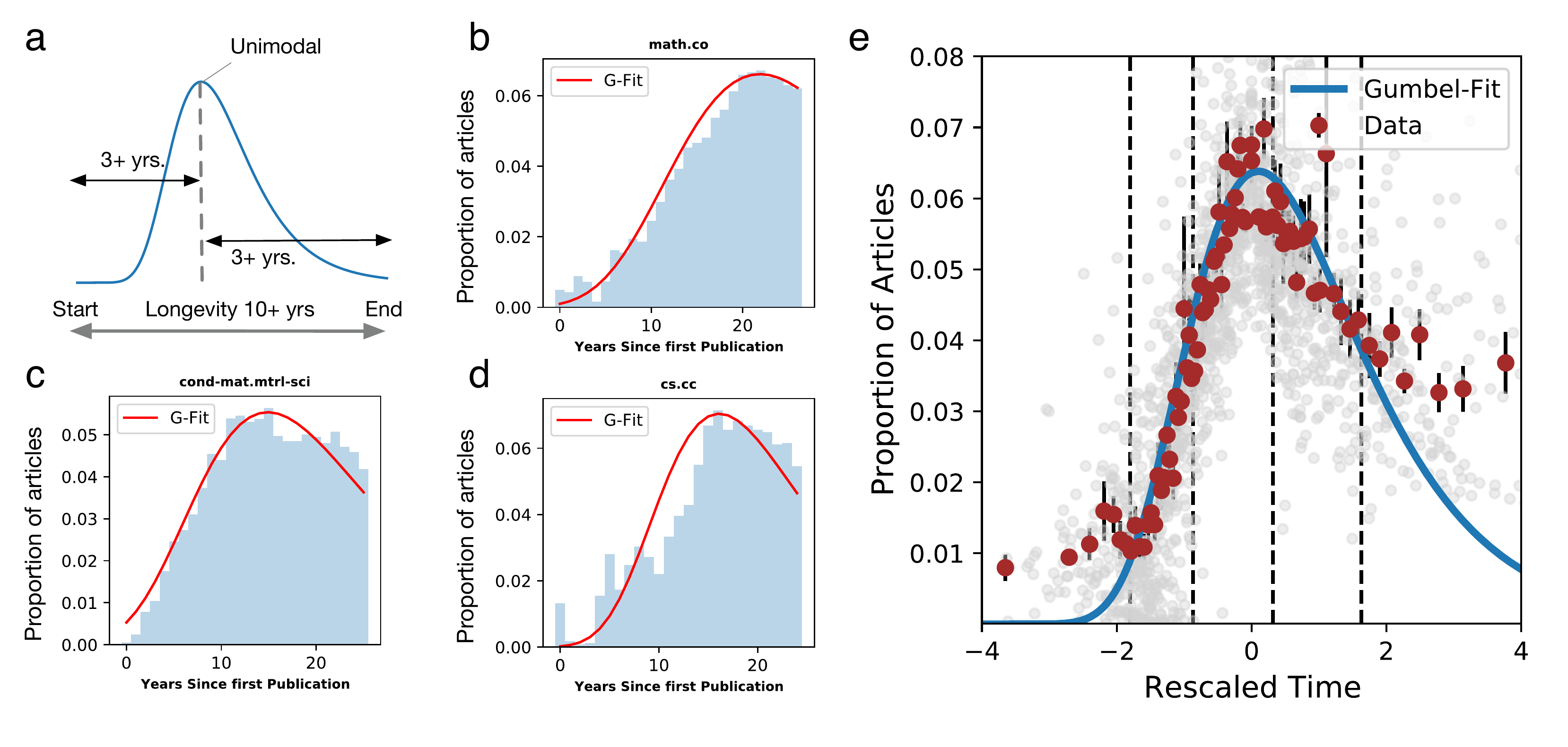}
    \caption{ \textbf{a}. Conditions for a field to be included in the analysis. \textbf{b-d} Gumbel fits for the fields with the largest numbers of articles in Physics, Mathematics and Computer Science: Mathematics - Combinatorics (b), Material Science (c) and Computational Complexity (d). \textbf{e}. Evolution of the 72 studied fields after temporal re-scaling from Eq. \ref{eq2}. The blue curve represents the Gumbel fit, and red dots correspond to the empirical average over equal-sized bins. Error bars indicate standard error.}
    \label{fig2}
\end{figure}

In order to estimate the model fit, we consider fields that satisfy three conditions: (i) longevity -- having at least 10 years of activity to ensure a sufficient observation period, (ii) unimodality -- we exclude multimodal distributions as it would require introducing a mixture model going beyond the scope of this study and (iii) completeness -- we require the peak of the distribution to be at least 3 years away from the beginning and the end of the collection period to ensure that we capture sufficient data on both sides of the distribution. This reduces the number of fields to $72$, which we consider in our analyses below.\\

Using a least-squares optimization fitting procedure (see Methods), we show that 66 out of 72 fields ($91.6\%$) exhibit a significant goodness of fit ($k<0.3$ and $p>0.05$ under KS-test, see Fig \ref{figS:fits_pval}). We show in Fig \ref{fig2}b-d the temporal distributions and Gumbel fits for the fields with the largest total numbers of articles in Physics, Mathematics and Computer Science. After obtaining the location $\alpha$ and scale $\beta$ parameters from the fitting procedure, we compute for each field the re-scaled time:

\begin{equation}\label{eq2}
    t^{'} = \frac{t-\alpha}{\beta}
\end{equation}

By re-normalizing fields with this standardized time, we observe that the various temporal distributions align on a single curve, highlighting the shared patterns of rise and fall across the fields studied (Fig \ref{fig2}e). In particular, the Gumble distribution provides a more stringent fit of the tails, as can be observed when comparing to a symmetric, Gaussian fit in a log scale (Fig \ref{figS:fits_pval}).

\subsection{Characterizing the stages of research field evolution}

Using the rescaled time from Eq. \ref{eq2}, we next explore the characteristics of articles and researchers at different stages of a research field evolution. We adopt hereafter the standard delineations of stages from the innovation diffusion literature \cite{rogers2010diffusion} and define 5 periods of research field evolution (creation, adoption, peak, early decay, and late decay) delineated at the re-scaled times corresponding respectively to the $2.5\%$, $16\%$, $50\%$ and $84\%$ quantiles of the Gumbel distribution in Fig\ref{fig2}e (blue curve). We then group articles within these categories for each field and examine the variation of their characteristics when averaging across all fields.\\ 

We consider characteristics of the articles submitted at various field stages, and of the authors who submit them. For articles, we focus on the number of fields reported (article multidisciplinarity), the number of authors (team size), the number of references made to other arXiv articles, and the number of citations received within arXiv (article impact). For authors, we consider their career stage at the time of submitting the article (seniority), the total number of articles submitted to arXiv (longevity), the number of fields their articles span during their career and the number of fields per article (author multidisciplinarity). We average these characteristics over the article coauthors for which we have a unique identifier (ORCID). In the case of career stage $s$, we use Eq. \ref{eq3}, where $N_{art}$ is the chronological rank of the current article across the author's publications and $N_{tot}$ is the total number of articles:

\begin{equation}\label{eq3}
    s = \frac{N_{art} - 1}{N_{tot}-1}
\end{equation}

\begin{figure}
    \centering
    \includegraphics[width=\textwidth]{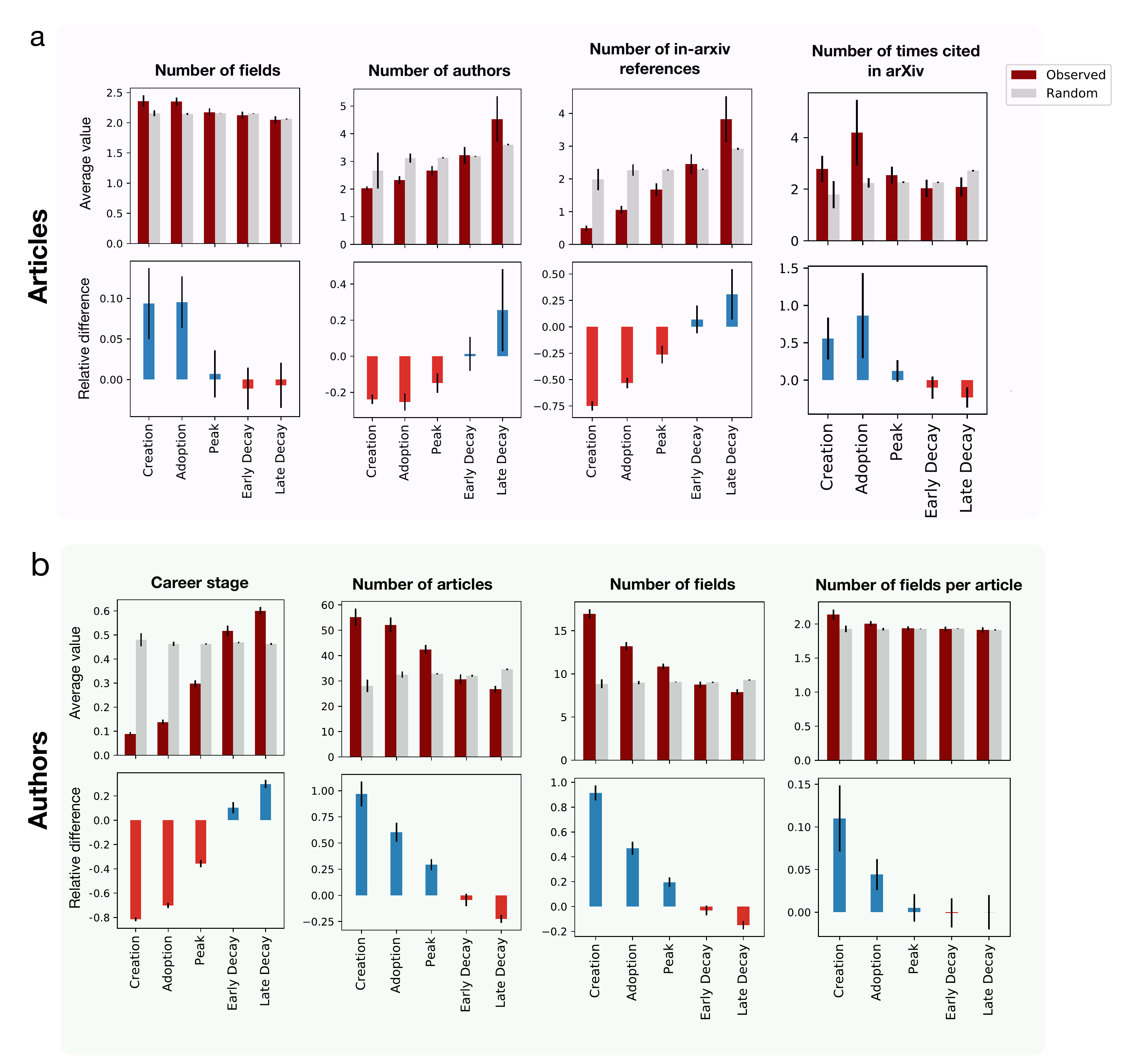}
    \caption{Characteristics of articles and authors at different evolutionary stages. The observed values are averaged over all fields (red bars). Gray bars correspond to the average field-specific random expectation (see Methods \ref{m4}). Bottom plots represent the relative difference between observed and random values. Error bars denote standard errors for observed values (red) and standard deviation for random values (gray). \textbf{a}  Article-centric features: number of fields reported in the article (multidisciplinarity), number of authors (team size), number of references made to other arXiv articles, and number of citations received within arXiv (impact).
     \textbf{b} Author centric features: career stage at the time of submitting the article (seniority), total number of articles submitted to arXiv (longevity), number of fields their articles span during their career and average number of fields per article (multidisciplinarity).
}
    \label{fig3}
\end{figure}

We show in Fig \ref{fig3} the average values of these features for each stage across the $72$ fields along with random expectations (see Methods \ref{m4}). In the context of article metrics (Fig \ref{fig3}a), we find that the early stages of research fields are characterized by interdisciplinary articles ($2.36$ fields, vs. $2.05$ for late decay) co-authored by small teams (2 authors vs 4.5). As fields evolve, we observe a steady growth in the number of references to earlier arXiv articles, indicating that the community builds on earlier works in arXiv (Fig. \ref{fig3} and Fig. \ref{figS:citations_in_field}a when restricting to the same field). Finally, we find that article impact, measured by the number of citations within arXiv, is maximal at the Adoption phase before the field has reached its peak. The citation count observes a similar trend in the case of total citations within arXiv shown in Fig. \ref{fig3}a as well as citations within arXiv received in the first five years (Fig. \ref{figS:citations_in_field}b). For author metrics (Fig. \ref{fig3}b), we find that the early stages of research fields are characterized by multidisciplinary authors (16.9 fields in career for creation vs 7.9 in career for late decay, and 2.13 fields per article vs 1.91 fields per article), who tend to be in their early career (8\% of total duration vs 60\%) with the longest longevity (55 papers vs 27).

\subsection{Cognitive distance and early innovation}

\begin{figure}[ht!]
    \centering
    \includegraphics[width=\textwidth]{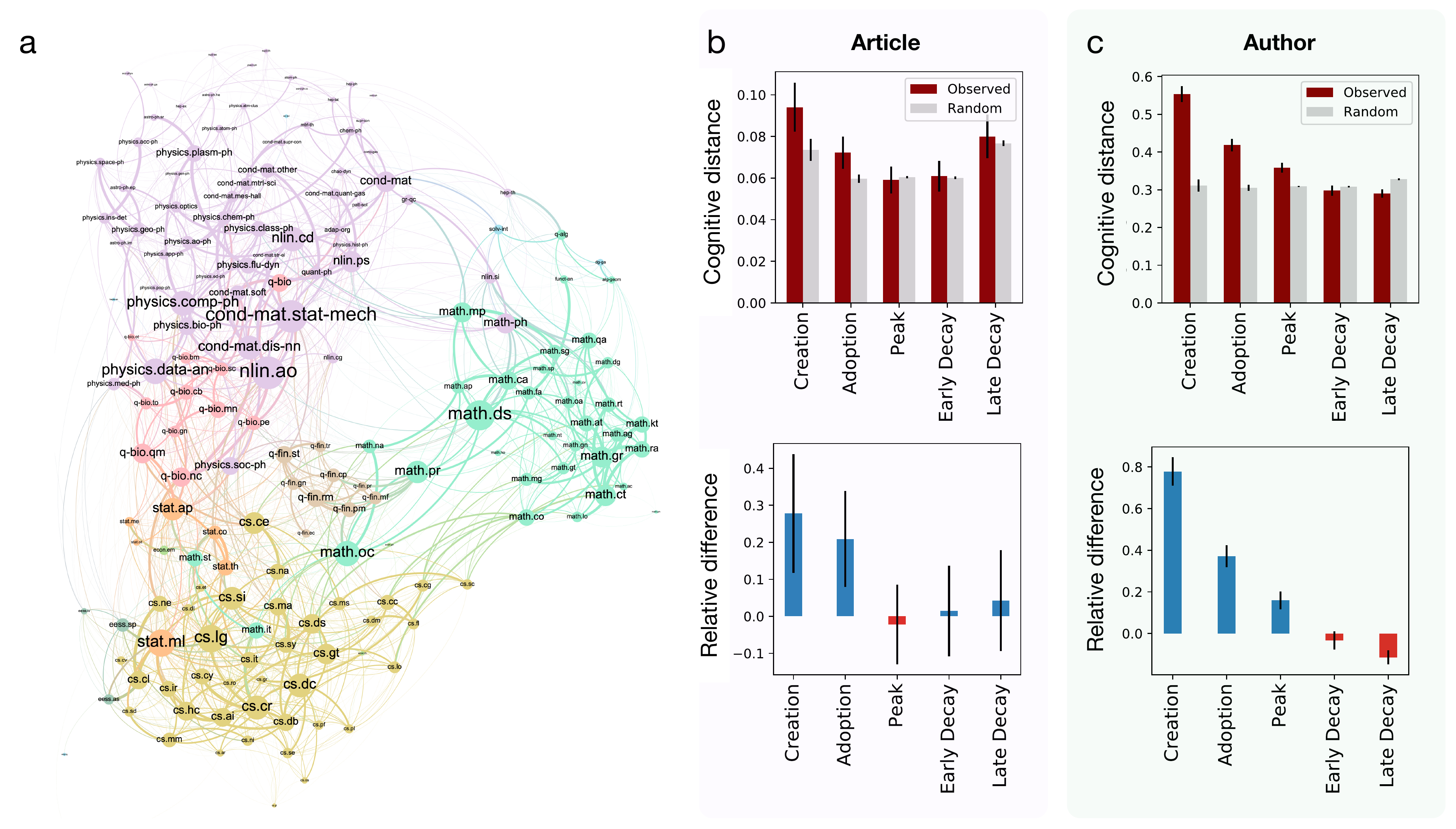}
    \caption{\textbf{a} Co-occurrence network of arXiv field tags. Nodes are colored based on the major research area they belong to (Physics, Computer Science, Mathematics, Statistics, Quantiative Finance, Quantitative Biology). Barplots in \textbf{b,c} follow the same method than in Fig \ref{fig3}. \textbf{b} Average cognitive distance across the field tags of articles. \textbf{c} Average cognitive distance across all the field tags used by authors throughout their career.}
    \label{fig4}
\end{figure}

The previous results show that works submitted in early phases of research fields tend to mix a larger number of field tags. However, this measure does not take into consideration the various levels of similarity between fields. For example, publishing an article within sub-fields of physics is different than publishing an article mixing quantitative biology, computer science, and physics. This is rendered apparent when examining the co-occurrence network of fields across arXiv articles (Fig. \ref{fig4}a). In the co-occurrence network, nodes represent field tags, and edges represent their co-occurrence across articles. To define edge weights, we first compute the number of co-occurrences between two fields across the whole period. We then compute a hypergeometric p-value that the two fields would have this number of co-occurrences given the number of times they each have occurred. Lower p-values indicate stronger similarity. We define the weight $W_{ij}$ between fields $i$ and $j$ as $-log_{10}(p_{ij})$, where $p_{ij}$ is the hypergeometric p-value. Edges with $p>0.01$ are finally filtered out. The network represents the landscape of fields in the arXiv, with closely related fields clustering together into communities corresponding to 6 broader categories: Physics (purple), Quantitative biology (gray), Computer Science (green), Mathematics (blue), Statistics (pink) and Quantitative Finance (orange).\\

Using this network embedding, we define the Cognitive distance $C_{i,j}$ between field tags $i$ and $j$ as the weighted shortest path $C_{i,j} = \sum_e \frac{1}{W_e}$, where $e$ are the edges on the shortest path between the two tags $i$ and $j$ and $W_e$ are their weights in the co-occurrence network. This cognitive distance allows us to provide a weighted proxy for interdisciplinarity. In particular, it allows to quantify the distance between disconnected fields: an example of this is shown in Fig. \ref{figS:cogdist_time}a where q-fin.ec (Economics) connects to hep-ph (High Energy Physics) by a path length of $4$.\\

We use this measure to compute for each article with at least two field tags the maximum cognitive distance between any pair of tags. We find that articles published in the early stages of a research field have a significantly larger cognitive distance, while the measure decays to the random level by peak stage (Fig \ref{fig4}b). Similarly, for authors we find that in earlier stages authors publish in cognitively distant fields, which narrows down to similar fields in later stages (Fig. \ref{fig4}c). The relative difference with random at the creation stage is more stringent that the previous measure using number of tags (articles: 0.3 vs 0.1, authors: 0.8 vs 0.1), strengthening our previous observation. 

\section{Discussion}

In this study, we leverage the field annotation of 1.5M articles from the arXiv preprint repository to explore the  scale-invariant patterns in the evolution of scientific fields and highlight the attributes of articles and researchers across different evolutionary stages. We show that research fields follow a right-tailed Gumbel temporal distribution, allowing to rescale their evolution over a single curve. We demonstrate the usefulness of this approach by highlighting characteristics shared by articles and authors across the various stages of a field evolution. We observe that early stages are characterized by articles written by small teams of early career, interdisciplinary authors, while late stages exhibit the role of large, more specialized teams. This supports the general finding that small teams disrupt while large teams develop science and technology \cite{milojevic2015quantifying,wu2019large}. We find that maximum impact, measured by citations, is reached before the peak of the field evolution during the Adoption stage. This may reflect foundational works underlying the subsequent attractivity of the field and moving it to the `peak' phase. In addition, we observe a steady increase in the within-field references to earlier work as fields evolve. This suggests a consolidation of the community over the particular body of work produced in the field, though further work on the citation and collaboration networks would be needed to investigate this aspect.\\ 

The main contribution of our work is to provide a method to rescale fields and associate research patterns to standardized evolutionary stages. However, this study has limitations. First, to capture sufficient data on the rise and fall patterns of research fields, we limited ourselves to a subset of 72 fields out of the 175 available. In particular, the choice of keeping only unimodal fields could be overcome by implementing a simple extension of our approach by using a mixture model, thereby capturing different ``waves'' of interest within a research field. In addition, to avoid ambiguity in author names we focused only on authors for which we could extract ORCID IDs, limiting the study to a small and potentially biased subset of authors. Future work should extend such analyses to larger databases with disambiguated authors and topic annotation to gain in generality.  Finally, while being an open repository, authors submitting to the arXiv need to be invited by another existing member from the main field of interest. These create social `chaperoning' constraints \cite{sekara2018chaperone} that might influence the type of authors observed at various stages.\\

Overall, this study contributes to the Science of Science literature by proposing a simple method to investigate the generic temporal properties of research fields, and highlighting its use in the context of arXiv. Future work should be conducted to provide mechanistic models recapitulating the observed patterns, and extending these analyses to larger datasets. We expect these insights to be helpful for researchers and policymakers interested in the emergence and development of research fields and more broadly in the dynamics of innovation \cite{ubaldi2021emergence}.

\section{Methods}

\subsection{Dataset extraction}\label{m1}
We extracted the publication metadata from the arXiv website using the arXiv API. The data spans years 1986 to 2018, with a total of 1,456,404 articles. For each article we retrieved the following characteristics: a) the unique article ID, b) the timestamp of article submission, c) the list of subjects categories (field tags), d) the citations received within arXiv, e) the references to other arXiv articles, and f) the list of last names of authors. We show an example article in Fig. \ref{fig1}a. Furthermore, we extracted when possible the ORCID IDs of the authors that declared it in arXiv. The number of unique ORCID IDs was 50,402, allowing to disambiguate these authors' names. 

\subsection{Fitting procedure}\label{m2}
\textbf{Uni-modality Test}: For filtering multi-modal fields we use the $diptest$ R library to compute the dip unimodality test. We remove fields that fail the test ($p<0.05$). \\

\textbf{Least square optimization}: For the selected fields, we strip years before the first publication to only consider years since first article. We then constrain the mode of the fitted distribution to coincide with the empirical one, and we fit the location and scale parameters using least-square optimization. 

\subsection{Assigning articles and authors to evolutionary stages}\label{m3}
We first collect for each field all articles containing the field tag. We associate each article to the evolutionary stage corresponding to the re-scaled time obtained for that particular field. We then assign the authors  of each article with an ORCID ID to the corresponding evolutionary stage. Note that articles with multiple field tags can be assigned to different stages of evolution corresponding to the re-scaled times of the different tags. 

\subsection{Randomization}\label{m4}

The observed  features in Fig \ref{fig3} are  compared  with  random  expectation  by  shuffling  for  each  field  the  re-scaled times across articles.  This procedure is repeated 50 times for each field and we compute the average for each stage.  Finally, we compute the average and standard deviation across fields.

\section{Acknowledgements}

Thanks to the Bettencourt Schueller Foundation long term partnership, this work was partly supported by the CRI Research Fellowship to Marc Santolini.

\printbibliography

\newpage

\appendix

\renewcommand\thefigure{S\arabic{figure}}    
\setcounter{figure}{0}    

\section{Supplementary Information}

\subsection{arXiv as a dataset}

Paul Ginsparg created arXiv in 1991. It was initially designed for sharing preprint articles with friends and colleagues \cite{ginsparg2011arxiv}. The reasons why researchers favor uploading their articles on arXiv are diverse. With a low threshold in the review phase and a minimal time between submission and online appearance, it provides a fast way for researchers to share their results with the scientific community. This in turn helps them in getting feedback from the larger ecosystem and gain intellectual precedence for their claims. The management team of arXiv follows a strict and systematic procedure ensuring accurate classification of an article to its subject domain (see Field tags management).
Though arXiv is lenient in its quality control as compared to a stricter ``peer-reviewed" system, an earlier study reports that $\sim 64\%$ arXiv articles end up publishing in WOS (Web Of Science) indexed journals and many journals also have started accepting arXiv preprint for submissions \cite{lariviere2014arxiv}, supporting the credibility of arXiv articles.\\ 

\textbf{Field tags management} - Users can choose appropriate field tags for their articles from the existing ones. They, however, cannot create their tags. The tags assigned by users are then reviewed by moderators of different subject domains and changed if deemed necessary. New field tags can only be introduced by the arXiv administration. They do consider proposals from researchers for introducing new tags and only after considering multiple factors such as the size of the research community, frequency of articles appearing in the field, or its impact on arXiv. A recent example of this was the introduction of two new tags in $2018$: econ.TH and econ.GN, corresponding to Economics Theory and Economics General. This happened after a community of economists proposed it to arXiv. However, most of the field tags appeared in the initial years (see Fig. \ref{figS:uniq_tags}). \\

\textbf{Growth rate} - To calculate the growth rate of the arXiv dataset, we consider the growth function as defined in Eq\ref{eq1}, with growth rate $r$:

\begin{equation}\label{eq1}
    N(t) = N_0e^{rt}
\end{equation}

\begin{figure}[ht!]

		\includegraphics[width=\textwidth]{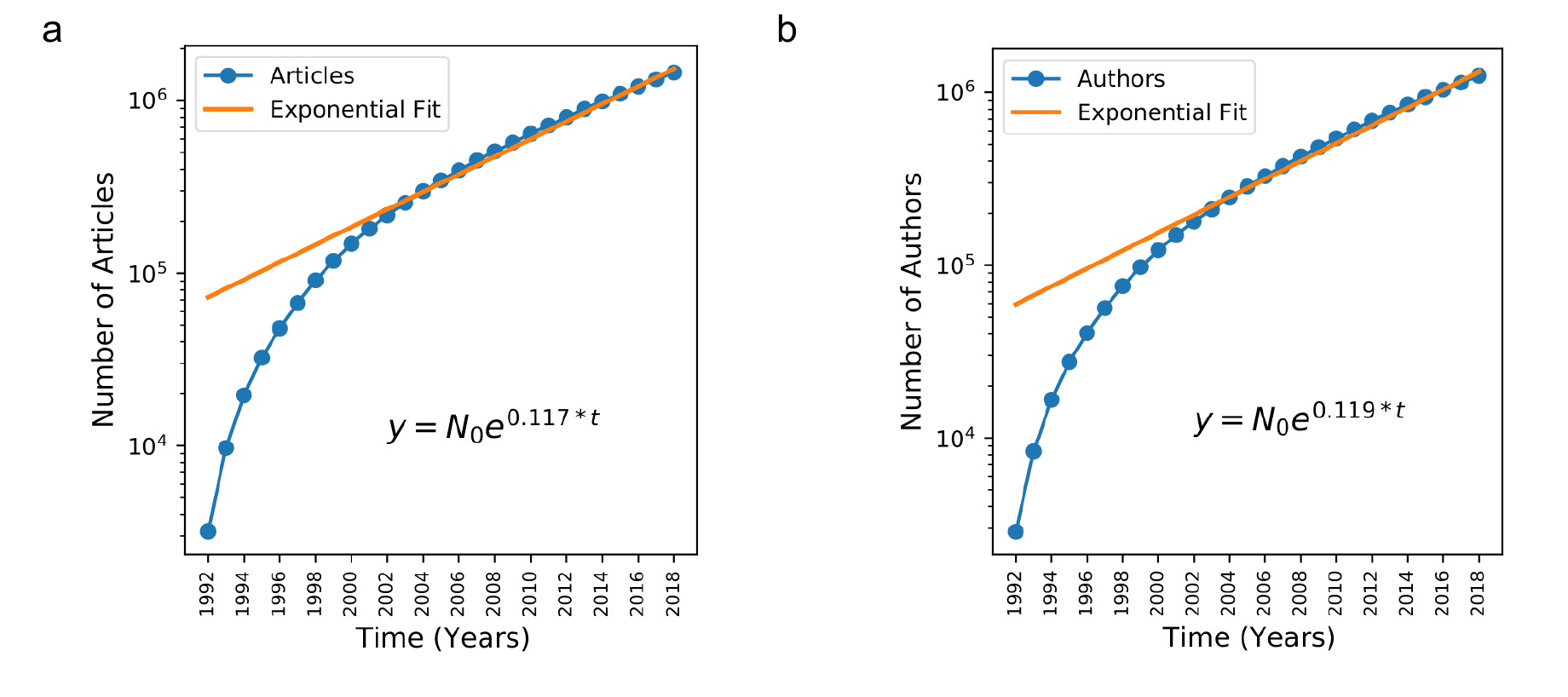}
		\label{fig1a}

	\caption{\textbf{a} Cumulative number of articles submitted to arXiv in time. \textbf{b} Cumulative number of (unique) authors. Both number of articles and of authors grow exponentially with a doubling period of $\sim$ 6 years.}
	\label{figS:num_authors}
\end{figure}

\begin{figure}[ht!]
    \centering
    \includegraphics[width=.7\textwidth]{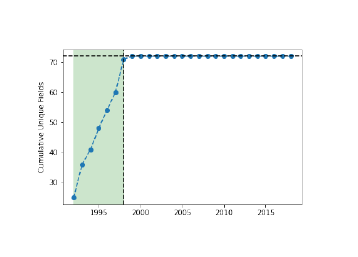}
    \caption{Cumulative number of field tags across time, among the $72$ studied. After an initial growth in the early years, the number of unique tags stays constant.}
    \label{figS:uniq_tags}
\end{figure}

We then fit the cumulative number of articles and number of authors in the dataset over time as shown in Fig. \ref{figS:num_authors}. 

The growth rates $r$ for articles and authors are respectively 0.117 and 0.119. Hence the doubling period i.e $\frac{ln2}{r}$ for articles and authors is resp. 5.9 and 5.8 years.

\subsection{Example of Gumbel distribution}

\begin{figure}[ht]
    \centering
    \includegraphics[width=0.9\textwidth]{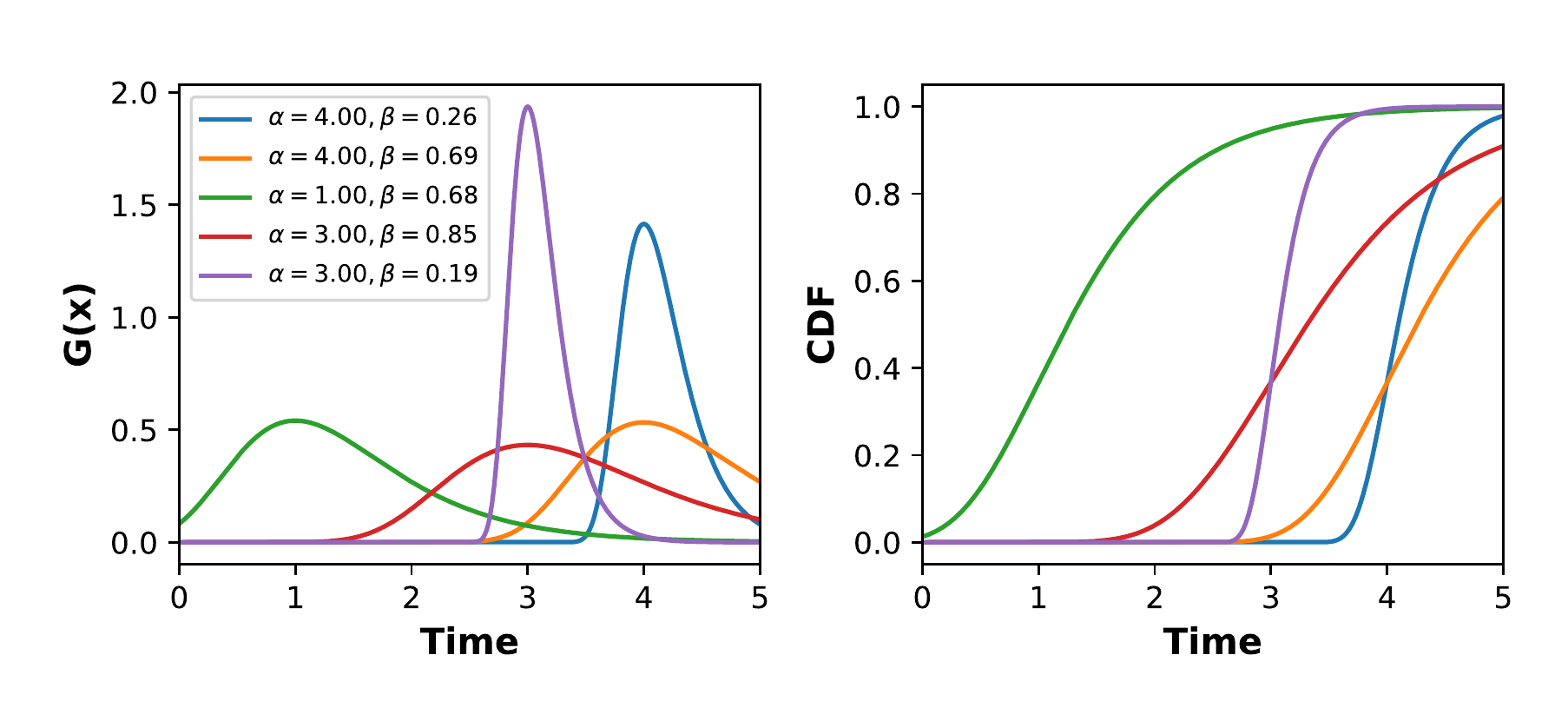}
    \caption{Examples of Gumbel distributions for different location and shape parameter values.}
    \label{figS:gumble_shapes}    
\end{figure}

\begin{figure}[ht]
    \centering
    \includegraphics[width=0.9\textwidth]{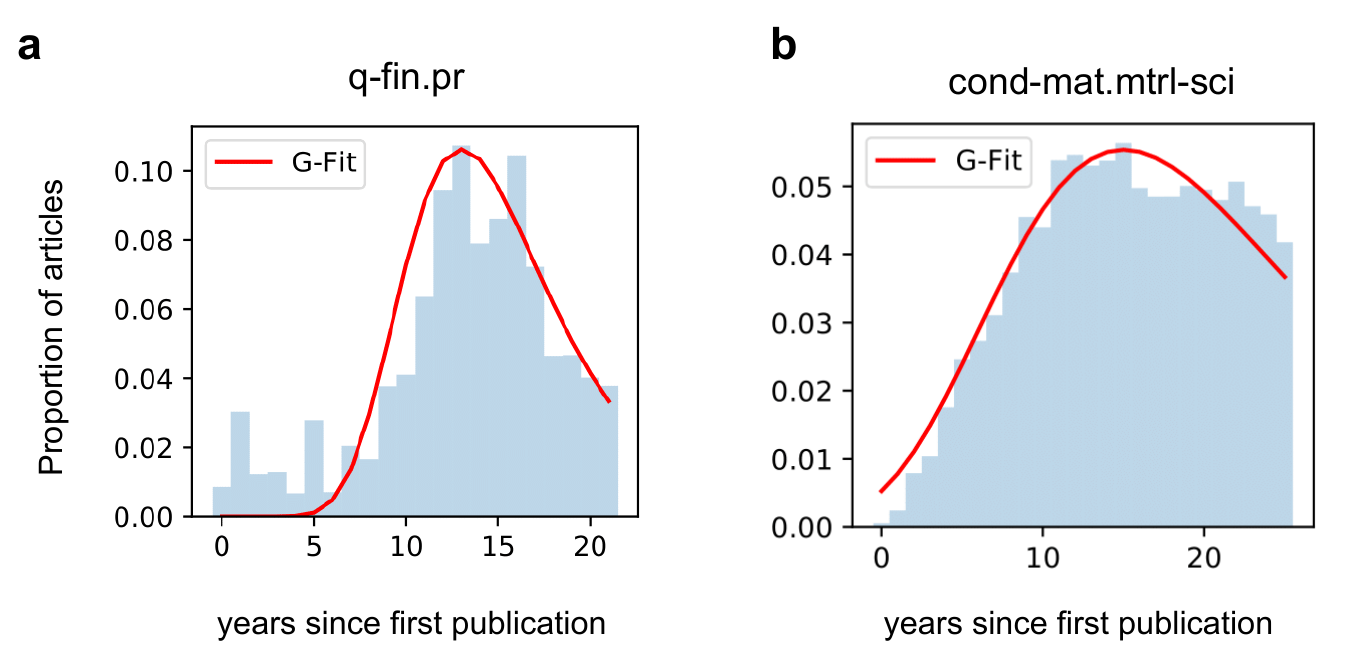}
    \caption{Empirical distribution and Gumbel fits for the fields of \textbf{a}  Quantitative Finance (q-fin.pr) and \textbf{b} Condensed matter material sciences (cond-mat.mtrl-sci).}
    \label{figS:example_gumbel}    
\end{figure}

To get more insights on the role of the $\alpha$ and $\beta$ parameters, we show in Fig. \ref{figS:gumble_shapes} some examples of Gumbel distributions with varying parameters. The location parameter $\alpha$ corresponds to the peak location, while the scale parameter $\beta$ corresponds to the distribution width. Fields with a low $\beta$ have a rapid rise followed by a rapid decay with a long tail. These could be the fields promoted by sudden advances in science and technologies or economics, for example, Pricing of Securities in Quantitative finance (q-fin.pr) (Fig. \ref{figS:example_gumbel}a). On the other hand fields with a large $\beta$ have a gradual rise and fall with a long tail in the decay phase -- for example Condensed Matter Material Sciences (cond-mat.mtrl-sci) (Fig. \ref{figS:example_gumbel}b).

\newpage
\subsection{Fitting the empirical data}

\textbf{Normalizing the Gumbel Distribution Function}

Since for each field we only observe a finite sampling period of the full distribution, we need to normalize the Gumbel distribution between times $t_1$ and $t_2$ to improve the fit. Given the Gumbel function $G(x)$, we find the normalizing constant such that:

\begin{align}
    C\int_{t_1}^{t_2} G(x,\alpha,\beta)dx &= 1 \\
    C\int_{t_1}^{t_2} \frac{1}{\beta}e^{\frac{-(x-\alpha)}{\beta}}e^{-e^{\frac{-(x-\alpha)}{\beta}}}dx &= 1 \label{eqGum}
\end{align}

Let $y = e^{-\frac{(x-\alpha)}{\beta}}$ $\implies$ 
$dy=-\frac{1}{\beta}e^{-\frac{(x-\alpha)}{\beta}}dx$. Replacing above in Eq.\ref{eqGum} and adjusting limits we get:

\begin{align}
    C\int_{e^{-\frac{(t_1-\alpha)}{\beta}}}^{e^{-\frac{(t_2-\alpha)}{\beta}}} -e^{-y} dy &= 1 \\
   C e^{-y} \Bigr|_{e^{-\frac{(t_1-\alpha)}{\beta}}}^{e^{-\frac{(t_2-\alpha)}{\beta}}} &= 1 \\
   C \left[ e^{-{e^{-\frac{(t_2-\alpha)}{\beta}}}} -    e^{-{e^{-\frac{(t_1-\alpha)}{\beta}}}} \right] &= 1\\
   C = \frac{1}{\left[ e^{-{e^{-\frac{(t_2-\alpha)}{\beta}}}} -    e^{-{e^{-\frac{(t_1-\alpha)}{\beta}}}} \right]} \label{eq6}
\end{align}

With the above $C$ value we can normalize the Gumbel distribution function for any values of $t_1$ and $t_2$. Note that when $t_1 \rightarrow -\infty$ and $t_2 \rightarrow \infty$ the constant $C \rightarrow 1 $.\\

\begin{figure}[ht!]
    \centering

        \centering
        \includegraphics[width=0.85\textwidth]{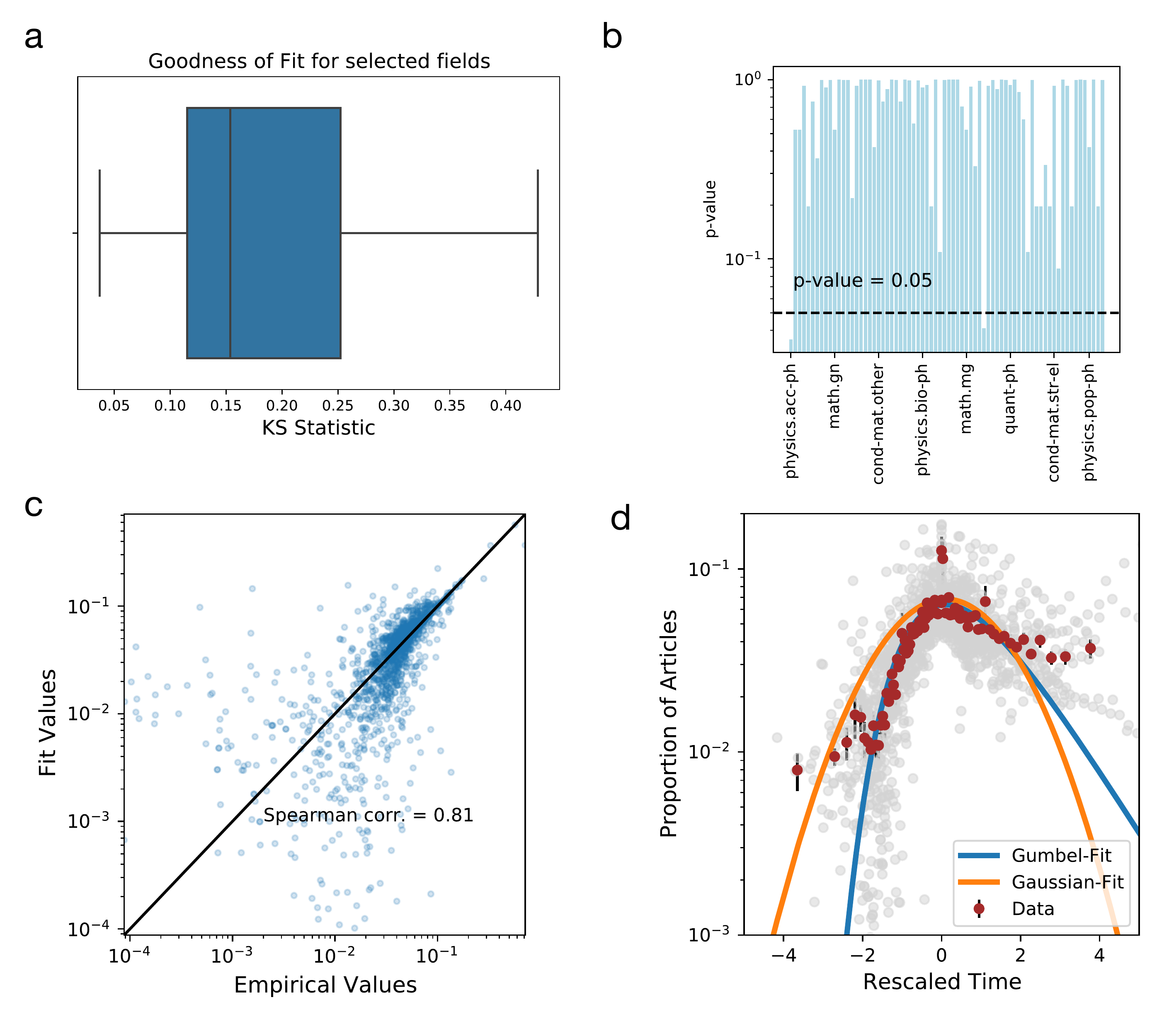}

    \caption{\textbf{a} Kolmogorov-Smironov test KS values for Gumbel distribution fits. Lower values indicate better fits.  \textbf{b} Corresponding p-values for the KS-test. Values of $p>0.05$ indicate a plausible fit. \textbf{c} Scatter plot of the fitted vs empirical values of the temporal distributions for the 72 selected fields. The Spearman correlation is $\rho=0.81$, with $p < 1e-16$. \textbf{d} Same as Fig \ref{fig2}e, in a log scale. We show both the Gumbel and Gaussian fits. The Gumbel fit provides a better description of the tails.}
    \label{figS:fits_pval}
\end{figure}

\begin{figure}[ht!]
    \centering
    \includegraphics[width=\textwidth]{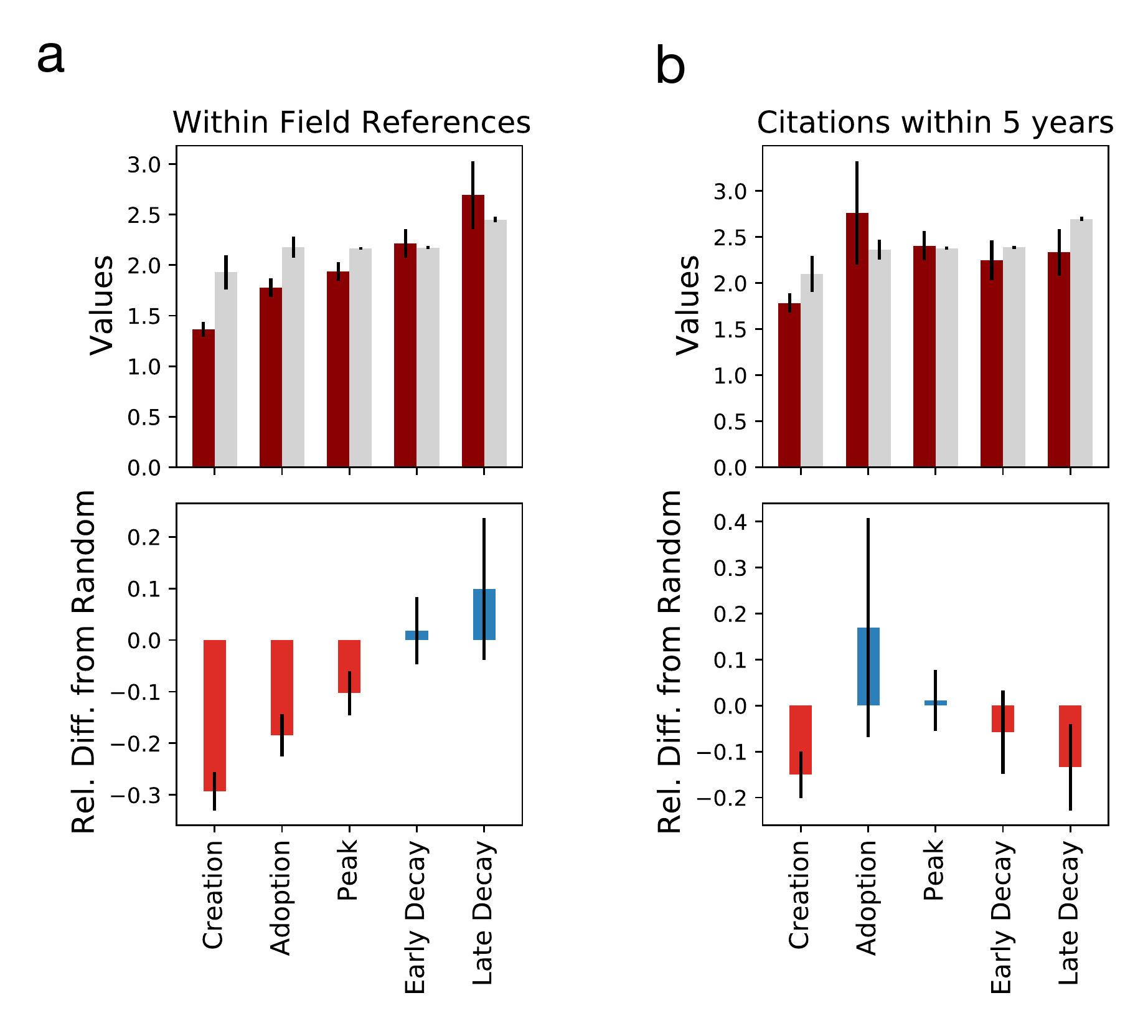}
    \caption{Same as Fig \ref{fig3}a, for references (\textbf{a}) and citations (\textbf{b}) within the same field than the article. Citations are limited to the 5 years following the article.}
    \label{figS:citations_in_field}
\end{figure}

\newpage

\begin{figure}[ht!]
    \centering
    \includegraphics[width=\textwidth]{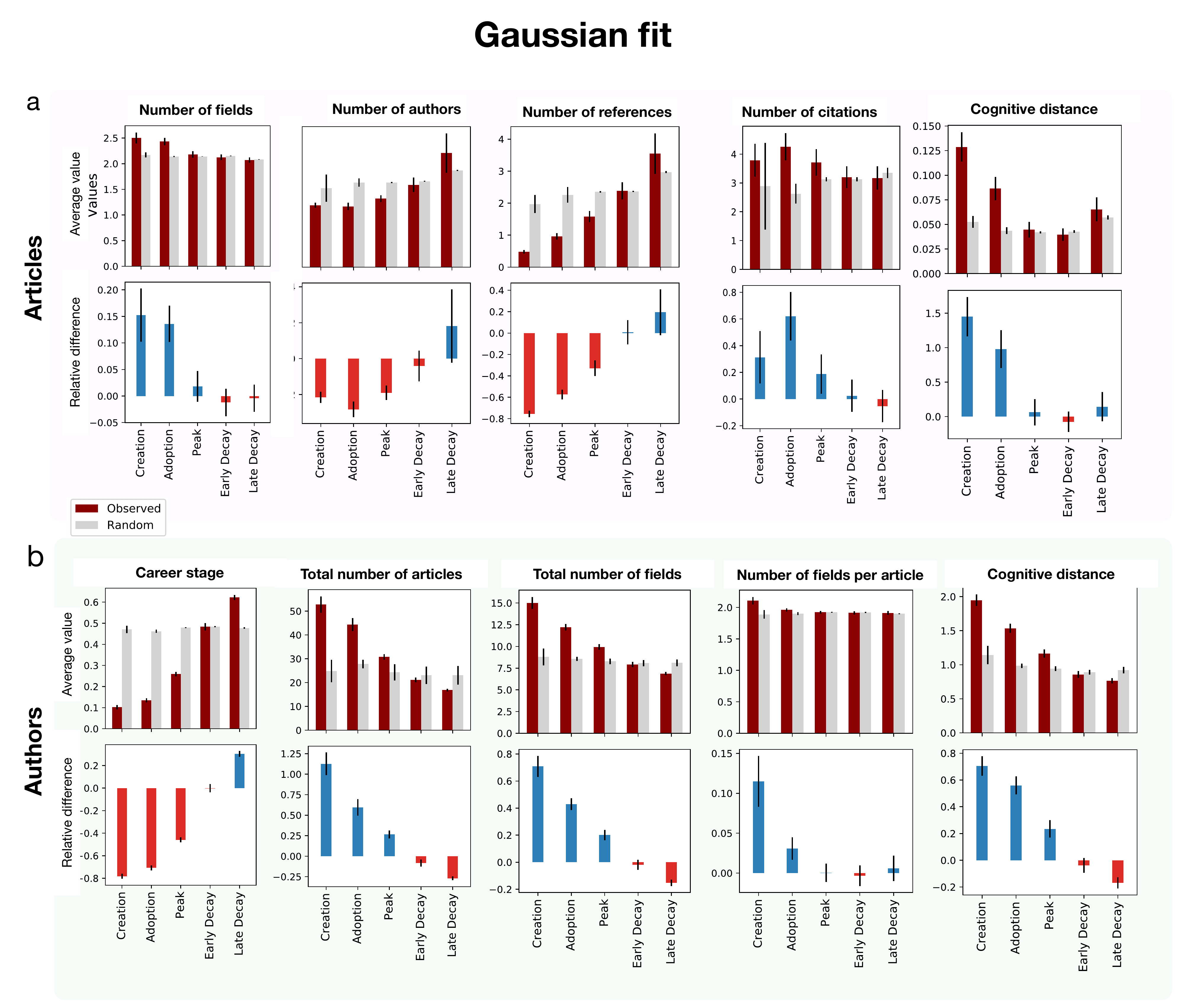}
    \caption{Article-centric and author-centric properties calculated with the Gaussian distribution fits.}
    \label{figS:features_gaussian}
\end{figure}

\begin{figure}[ht!]
    \centering
    \includegraphics[width=\textwidth]{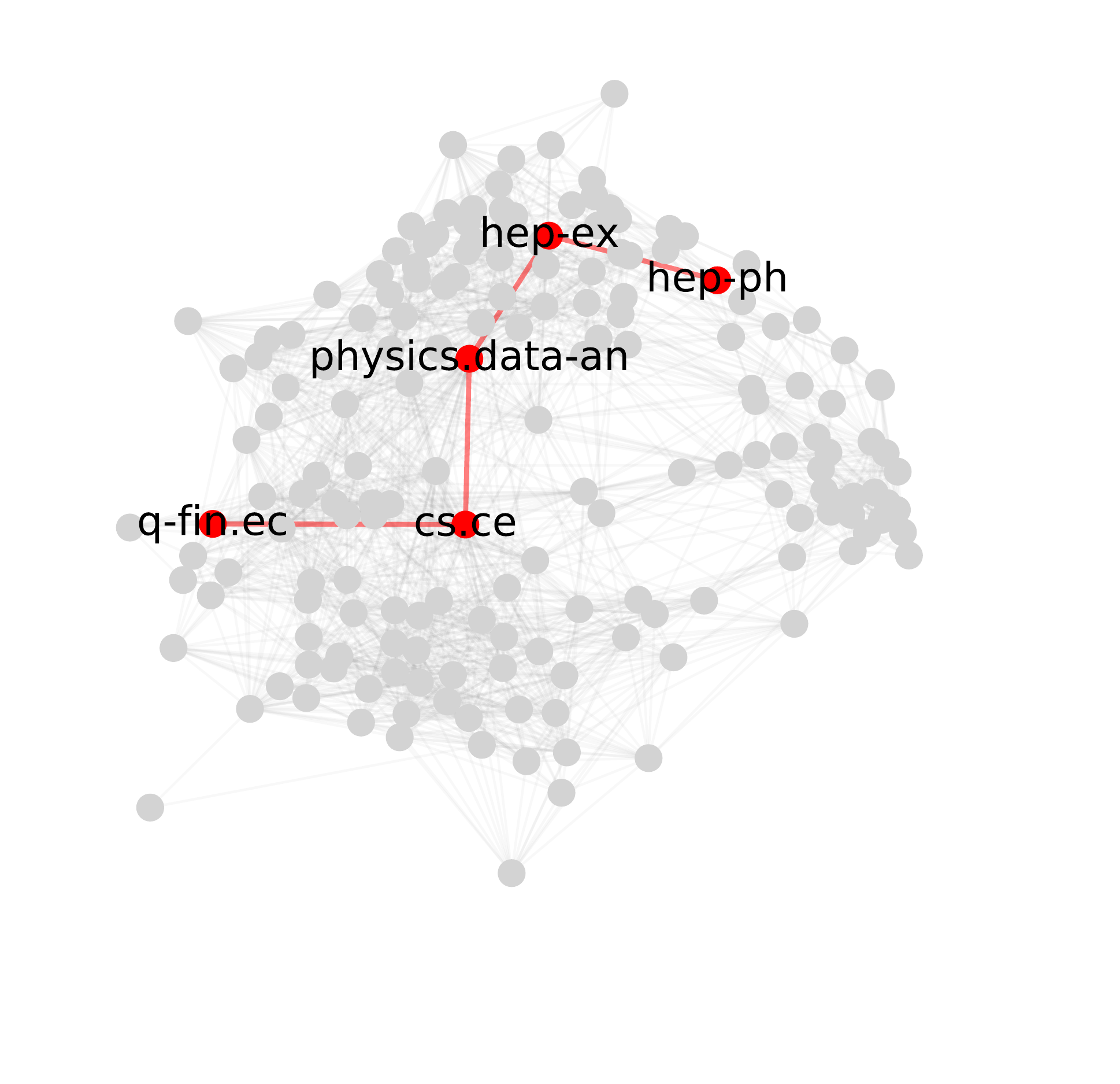}
    \caption{ Example of a shortest path linking the distant fields of Quantitative finance and High Energy Physics in the field co-occurrence network.
    }
    \label{figS:cogdist_time}
\end{figure}

\end{document}